\documentclass [floatfix, superscriptaddress, twocolumn, showpacs, aps, pra, 10pt]{revtex4-1}
\usepackage{amsmath}
\usepackage{natbib}
\usepackage{hyperref}
\usepackage{graphicx}
\usepackage{bbm}
\usepackage{amssymb}
\usepackage[english]{babel}
\usepackage{lmodern}
\usepackage[T1]{fontenc}
\usepackage{wasysym}
\usepackage[normalem]{ulem}
\usepackage{microtype}
\usepackage[per-mode=symbol]{siunitx}
\DeclareSIUnit\gauss{G}
\usepackage{booktabs}

\usepackage{dcolumn}
\usepackage{color}

\graphicspath{{Figs/}}

\usepackage{amssymb}

\usepackage{ulem}

\begin{document}

\title{Electromagnetic Induction Imaging: Signal Detection Based on Tuned-Dressed Optical Magnetometry}

\author{Giuseppe Bevilacqua}
\author{Valerio Biancalana}
\email{valerio.biancalana@unisi.it}
\affiliation{Dept. of Information Engineering and Mathematics - DIISM, University of Siena -- Via Roma 56, 53100 Siena, Italy}
\author{Yordanka Dancheva}
\affiliation{Dept. of Physical Sciences, Earth and Environment - DSFTA, University of Siena -- Via Roma 56, 53100 Siena, Italy}
\affiliation{currently with Aerospazoio Tecnologie, Rapolano (SI), Italy}
\author{Alessandro Fregosi}
\affiliation{Dept. of Physical Sciences, Earth and Environment - DSFTA, University of Siena -- Via Roma 56, 53100 Siena, Italy}
\affiliation{CNR Istituto Nazionale di Ottica, via Moruzzi 1, 56124 Pisa, Italy}

\author {Gaetano Napoli}
\affiliation{Dipartimento di Matematica e Fisica "Ennio De Giorgi"  Universit\`a del Salento, Lecce, Italy
}
\author{Antonio Vigilante}
\affiliation{Dept. of Physical Sciences, Earth and Environment - DSFTA, University of Siena -- Via Roma 56, 53100 Siena, Italy}

\begin{abstract}
 A recently introduced tuning-dressed scheme makes a Bell and Bloom magnetometer suited to detect weak variations of a radio-frequency (RF) magnetic field.  We envisage the application of such innovative detection scheme as an alternative (or rather as a complement) to RF atomic magnetometers in electromagnetic-induction-imaging apparatuses.  
\end{abstract}

\date{\today}

\maketitle

\section{Introduction}
\label{sec:introduction}

In 2001, Griffiths  \cite{griffiths_mst_01} proposed an imaging technique based on inferring one or more of  the three passive electromagnetic properties (conductivity $\sigma$, permittivity $\epsilon$ and permeability $\mu$) to produce images on the basis of the response to a position dependent oscillating magnetic field. Several denominations are used to identify this kind of methodologies, among which electromagnetic induction imaging (EII)  \cite{deans_apl_20}, electro-magnetic tomograhy (EMT) \cite{yu_el_93}, magnetic induction tomography (MIT) \cite{Korjenevsky_pm_00, bevington_as_20}, and also mutual inductance tomography (same acronym) \cite{soleimani_ieee_07}, the  latter three stressing  the  potential of the technique to provide 3D mapping. A review on the subject was recently authored by Ma and Soleimani \cite{ma_mst_17}. 

As a general feature, the technique uses an AC magnetic field to excite eddy currents in the specimen. The secondary magnetic field generated by  those currents is then detected and analyzed.  

Magnetometers can detect and record in-phase ($\epsilon$ dependent) and out-of-phase ($\sigma$ dependent) response of specimens subjected to an excitation radio-frequency field $B_{\mathrm{rf}}$. The response obtained while scanning the  position of a sample with respect to the field generator and detector enables the registration of 2D maps. The possibility of varying the $B_{\mathrm{rf}}$ frequency (and correspondingly the skin depth in the specimen) enables 3D (tomographic) capabilities.

Optical magnetometers, in the so-called radio-frequency (RF) implementation \cite  {deans_apl_16,wickenbrock_apl_16} are excellent detectors of weak (variation of) magnetic fields oscillating at a resonant frequency and hence constitute favourite detectors for EII, particularly in the case of low-conductivity and/or small size specimens.

This led recently to important steps toward high resolution EII of weakly conductive materials. Appositely  developed RF magnetometers with opportune field-specimen arrangement, \cite{jensen_prr_19,marmugi_apl_19},  demonstrated a sensitivity sufficient to detect and characterize sub S m$^{-1}$ conductive material in small size (\SI{5}{\milli l}) specimens, despite the operation in unshielded environment) \cite{deans_apl_20}. 

Such specification level makes the technique suited to develop tools for medical diagnostics \cite{scharfetter_ieee_03}, where the negligible invasivity of EII represents a valuable attractive, but the low-conductivity  and the need of high-spatial resolution constitute a severe requirement.

This work considers the potential of a peculiar configuration of a Bell\&Bloom (BB) magnetometer as a detector of eddy currents induced in small and/or weakly conductive specimens, which could be eventually used to produce maps  in 2D or 3D scans.

While the basic principle of operation (detection of a secondary magnetic field) is shared with the above mentioned research, the proposed sensor works with a particular arrangement that does not require resonant conditions for the time-dependent excitation field. The proposed arrangement can constitute  an alternative as well as a complement to the commonly used RF magnetometers.

The standard BB implementation of atomic magnetometers is based on optically pumping atoms into a given Zeeman sublevel (typically with a maximum magnetic number along the quantization axis defined by the pump light wavevector) that is not stationary due to the presence of a transverse magnetic field. 

In the usual picture of the BB operation, the macroscopic magnetization precesses around the magnetic field direction and it is periodically reinforced when its orientation is along the optical axis of the pump beam. The periodic reinforcement is obtained by  modulating the pump radiation --its intensity, polarization or wavelength-- synchronously with the precession. A weak and unmodulated probe radiation interrogates the evolution of the atomic state and produces an output signal whose dynamics is driven by the field under measurement. 

The presence of time dependent field can modify the spin dynamics also in non-resonant conditions. As an example, in  conventional BB magnetometry, slow-varying   magnetic field variations, oriented along the bias field direction,  are detected with a high sensitivity. Field variations along perpendicular directions produce some signal, as well, but with a second-order response \cite{biancalana_arx_21}.

Interesting dynamic responses of precessing spins are observed in a different regime, namely in the case of an intense fast-varying magnetic field.  In particular, a strong magnetic field oscillating along the probe-beam axis (i.e. transversely to the static one) at a frequency much above the resonance may effectively freeze the atomic precession, according with a phenomenon originally studied in the late Sixties \cite{haroche_prl_1970} and generally known as \textit{magnetic dressing}.   .

In a recent work \cite{bevilacqua_prl_20}, we have demonstrated that the dressing phenomenon can be deeply tuned by the application of a secondary (and much weaker) field that oscillates at the same frequency --or at a low-order harmonic-- of the dressing one, perpendicularly to the latter. Among the peculiarities of such tuned-dressed configuration, we  pointed out that the system shows a remarkable response to small variations of the amplitude and/or of the phase of the tuning field. This work investigates with a proof-of-principle experiment the potential of such a tuning-dressing apparatus to detect tiny field variations due to eddy currents induced by the tuning field.

\section{Experimental setup}
\label{sec:setup}
The experimental setup is built around a BB magnetometer making use of cesium thermal vapor in a centimetric, gas buffered cell. The atomic vapour is synchronously pumped by $D_1$ radiation (at \SI{}{\milli\watt\per\square\centi\meter} irradiance) and polarimetrically probed by a weak (at \SI{}{\micro\watt\per\square\centi\meter} irradiance) $D_2$ radiation, co-propagating with the pump one along the $x$ direction, further  details  can be found in Ref. \cite{biancalana_apb_16}.
 
The atomic sample is merged in a  constant field at $\mu$T level (let its direction be $z$) oriented perpendicularly to the laser beams. This static field $B_0$ is obtained by partially compensating the environmental magnetic field. The task is accomplished by means of three large size (180 cm) mutually orthogonal Helmholtz pairs. Additional adjustable quadrupoles let improve the field homogeneity.

\begin{figure}
    \centering
    \includegraphics[width=0.8 \columnwidth]{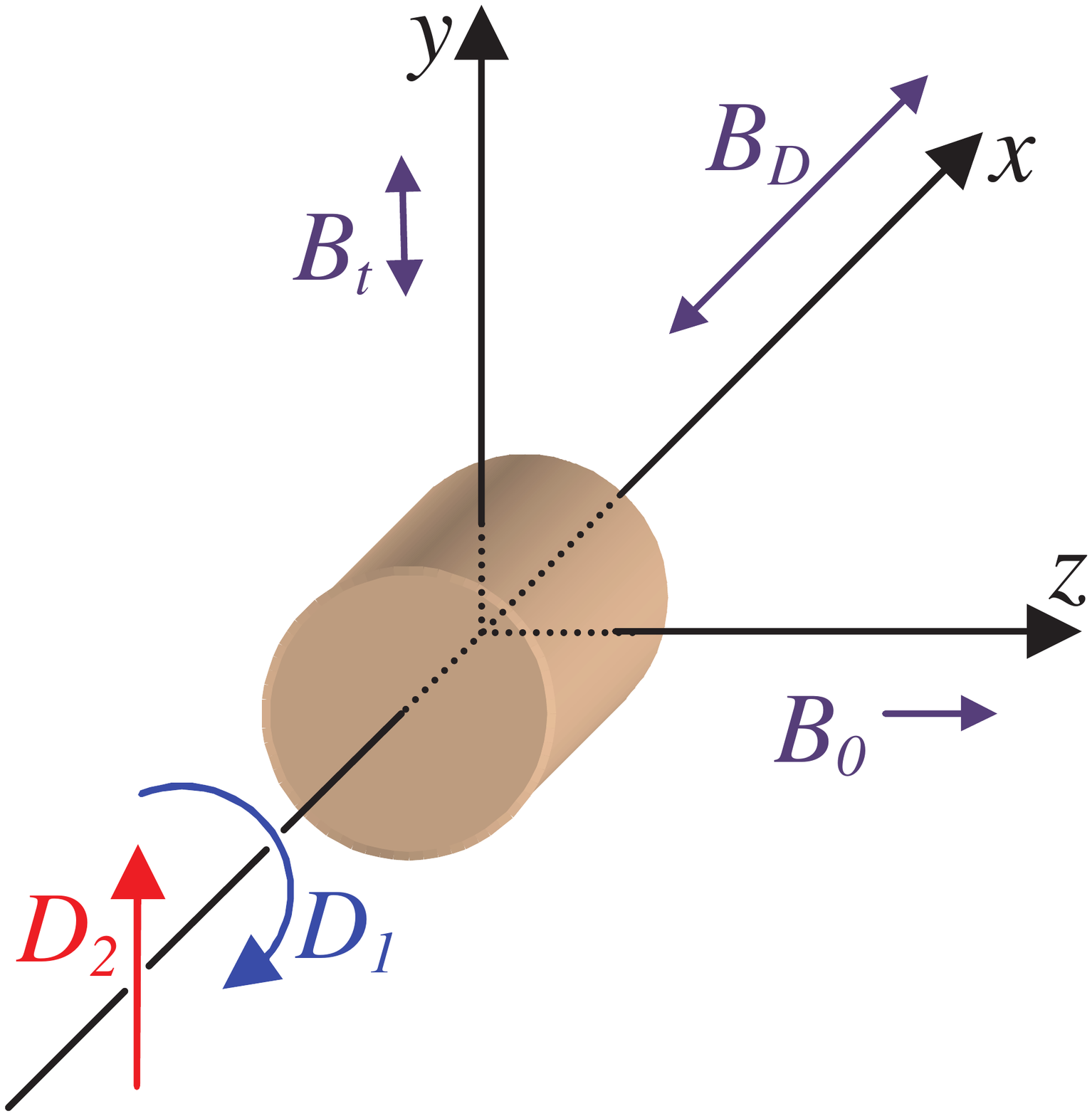}
    \caption{Magnetic field and laser beams geometry. Circularly polarized $D_1$ pump radiation and linearly polarized $D_2$ probe radiation propagate along $x$, which is the same direction of a strong oscillating (dressing) field $B_D$. A weaker (tuning) field $B_t$ oscillates along $y$ synchronously with $B_D$; a weak static field $B_0$ is oriented along $z$.}
    \label{fig:geometria}
\end{figure}

 The tuning-dressing effect \cite{bevilacqua_prl_20} occurs when two phase-related oscillating fields ($B_D$ and $ B_t$) are applied along $x$ to dress the atoms, and along $y$ to tune the dressing effect, respectively. 
 A solenoidal coil (\SI{11}{cm}  in length and \SI{4}{cm} in diameter) surrounding the cell generates $B_D$ and a small (millimetric) solenoid wound on a ferrite nucleus generates $B_t$.
 The interaction geometry is sketched in Fig.\ref{fig:geometria}.

The $B_D$ and $B_t$ coils are supplied by two waveform generators (Agilent 33250A) phase-locked to each other. Series capacitors help to adapt the impedances, and a linear amplifier can be used to enhance the $B_t$ amplitude. Each of these coils has \SI{10}{Ohm}   series resistor to precisely monitor the current phases and amplitudes via a \SI{16}{bit}, \SI{500}{\kilo S/s}   DAQ card (NI 6346).

The tuning-dressed phenomenon occurs and is well modeled and characterized for a tuning-field frequency $\omega_t$ that is an integer multiple of the dressing frequency $\omega_D$  ($\omega_t=p \omega_D$). It is worth stressing that, in the considered application, switching among different $p$ values may enable fast and relevant variations of the skin depth, which in EII applications may constitute an interesting feature.

\begin{figure}
    \centering
    \includegraphics[width=0.8 \columnwidth]{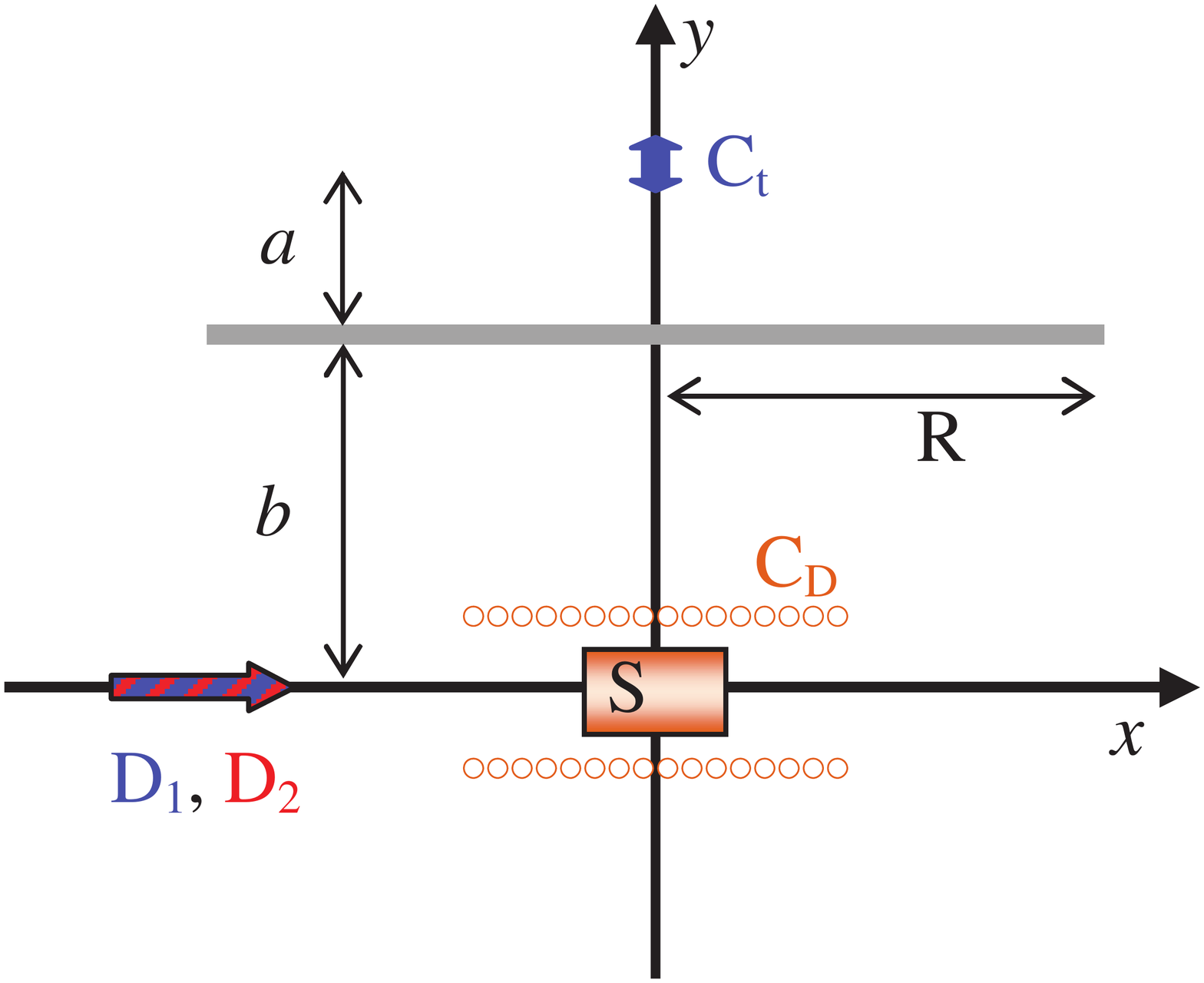}
    \caption{Relative positions of specimen, sensor, and tuning/dressing field sources. The specimen is a circular disk of radius R (thick grey line), which is parallel to the $xz$ plane and is centered on the $y$ axis, at a quote $y=b$ over the sensor $S$. A solenoidal coil $C_D$ surrounds $S$ and produces a homogeneous field $B_D$ along the $x$ axis. The latter coincides with the optical axis of the pump and probe lasers $D_1$ and $D_2$. A small dipolar source $C_t$ located on the y axis at distance $a$ from the specimen (and $a+b$ from the sensor) produces a tuning field $B_t$ which is oriented along $y$ on the sensor, and excites tangential eddy currents in the disk. Those currents modify both the amplitude and the phase of the tuning field on $S$.}
    \label{fig:interazione}
\end{figure}

The coil-specimen-sensor geometry is  sketched in Fig.\ref{fig:interazione}. The used specimens are one or more Aluminum disks of assigned thickness $d$, diverse radii $R$, centered on the $y$ axis, and lying on a $xz$ plane between the $B_t$ generator and the atomic sensor. The experimental results reported in this work are obtained with $a=$\SI{40}{mm}, $b=$\SI{130}{mm}, $d=$\SI{30}{\micro m},
and $D$ ranging from  \SI{24}{mm} to \SI{350}{mm}. In this proof-of-concept experiment, these values are chosen on the basis of fortuitous constraints of the available setup. In real imaging applications, all of them could be selected in view of optimizing the detection performance, in terms of sensitivity and/or spatial resolution.

The eddy currents induced in the conductive disks modify the amplitude and the phase of $B_t$ in the sensor location. As summarized in Sec.\ref{sec:model}, both these parameters play a role in shifting the effective Larmor frequency, and this is the key feature at the basis of the proposed detection technique.

\section{Effective Larmor Frequency \label{sec:model}}
As shown and discussed in the Ref.\cite{bevilacqua_prl_20}, a magnetic field that oscillates in the direction perpendicular to both the transverse static field $B_0$ and the longitudinal dressing field $B_D$ at a frequency $p$ times larger than that of the dressing field, modifies the effective Larmor frequency according to

\begin{equation}
\label{eq:poddmodel}
\Omega_L=\gamma \left[ B_0 J_0(\xi)+  B_t J_p(\xi) \sin(\phi) \right],
\end{equation}
if $p$ is an odd integer, and to
\begin{equation}
\label{eq:peven}
\Omega_L=\gamma\sqrt{\left[ B_0 J_0(\xi)\right]^2+ \left[ B_t J_p(\xi) \cos(\phi)\right]^2},
\end{equation}
if it is an even one. 
Here, $\gamma$ is the atomic gyromagnetic factor ($\gamma \approx $\SI{3.5}{\hertz\per\nano\tesla} for Cesium), and the argument $\xi$ is set by the amplitude of the dressing  field $B_D$ and by its frequency $\omega_D$ according to 
\begin{equation}
\xi=\frac {\gamma B_D}{\omega_D} .    
\end{equation}
 
The validity of the  eqs.\eqref{eq:poddmodel} and \eqref{eq:peven} requires that 
\begin{equation*}
    \xi \gg \frac{\gamma B_t}{\omega_D}, \frac{\omega_0}{\omega_D}
\end{equation*}

In the measurements shown in this paper, the tuning field oscillates at the same frequency of $B_D$ (i.e. $p=1$), while both its amplitude and  phase vary either due to different waveform-generator settings or to the presence of conductive samples placed in the proximity of the field source and of the sensor. In particular assuming that $\phi=\varphi-\theta$, where 
$\varphi$ is the relative phase of the two RF generators, and  $\theta$ is the dephasing caused by the eddy currents in the specimen, the effective Larmor frequency can be expressed as 
\begin{equation}
\label{eq:fdiathetaphi}
f=\frac{\gamma}{2 \pi}\left[J_0(\xi) B_0+ J_1(\xi) B_t \sin{\phi}\right ]=f_0 + A \sin(\varphi-\theta),
\end{equation}
where $f_0$ is the dressed frequency in absence of the tuning field. Beside causing the dephasing $\theta$, the eddy currents attenuate $B_t$, and this shielding effect determines a reduction of the parameter $A$.

The eq.\eqref{eq:fdiathetaphi} lets define  the conditions in which  the system response to either the attenuation or the dephasing is maximized:
\begin{equation}
\label{eq:dfdA}
\left| \frac{\partial f}{\partial A} \right | =\left | \sin(\varphi-\theta) \right |
\end{equation} is maximal for $\phi=\varphi-\theta=\pi/2  $, while
\begin{equation}
\label{eq:dfdtheta}
\left| \frac{\partial f}{\partial \theta} \right |= \left |A \cos(\varphi-\theta) \right |
\end{equation} is maximal when $J_1$ is maximum, that is at $\xi \approx 1.84$ and $\phi= 0, \pi$. These relations provide an indication of good experimental working conditions.

\section{Results}
\label{sec:results}

\begin{figure}
    \centering
    \includegraphics[width=0.9\columnwidth]{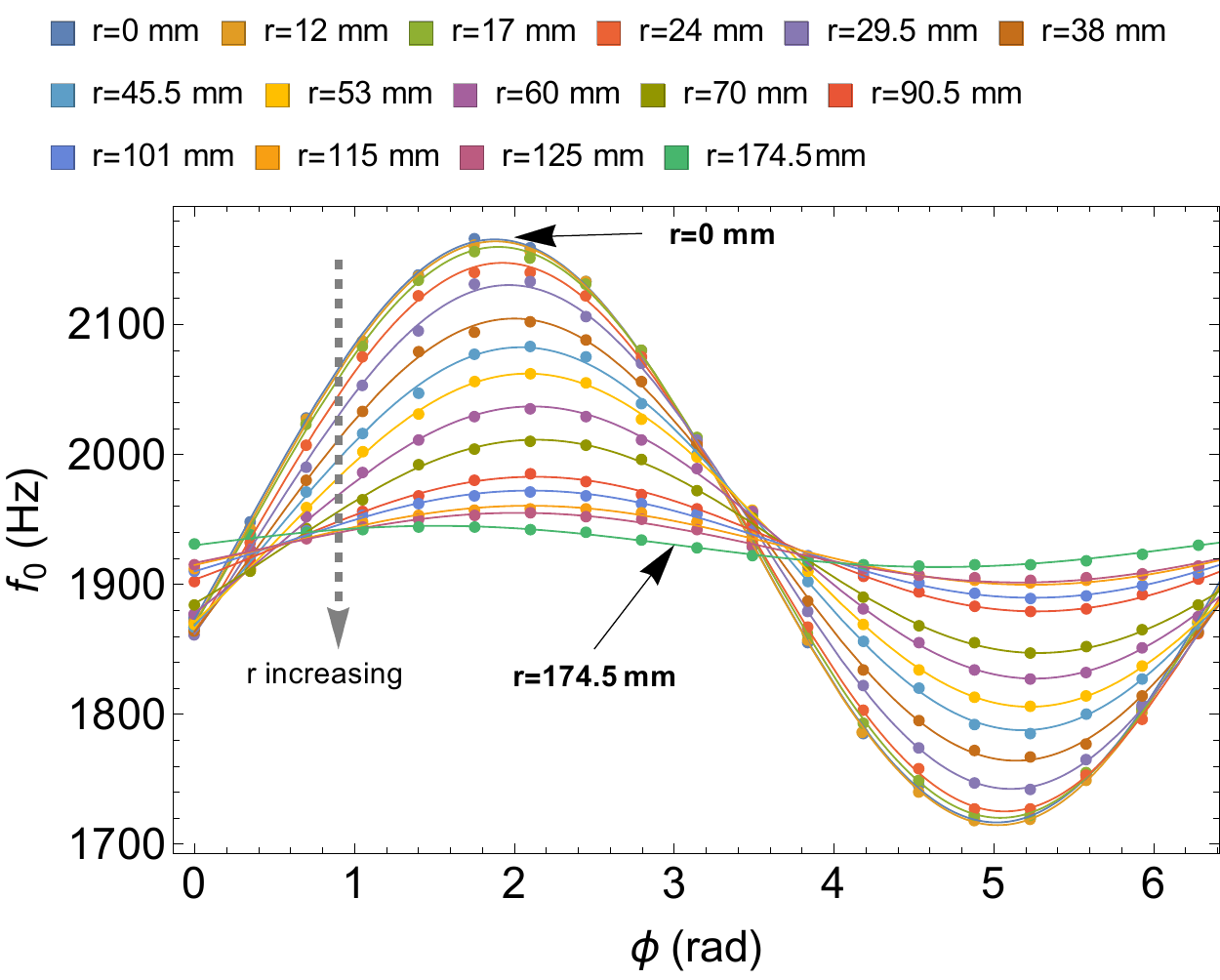}
    \caption{The resonance frequency as a function of the different relative phase $\varphi$, for various disk diameters whose size in mm is reported in the legend. The vertical error bar ($\approx\SI{3}{\hertz}$) is smaller than the symbols used in the graph. The solid lines represent the best fit profile obtained with the target function eq\eqref{eq:fdiathetaphi}, with $A$ and $\theta$ as fitting parameters.
    }
    \label{fig:sinusoidi}
\end{figure}

The $B_t$ variation due to the eddy currents induced in the specimen can be evaluated analytically for large (infinite radius) disks \cite{dodd_jap_68}, or numerically for finite radius ones. 
The variation is monotonically dependent on the disk radius, and concerns both the phase and the amplitude of $B_t$. This is verified experimentally as shown in Fig.\ref{fig:sinusoidi}

The atomic resonance is experimentally analyzed by scanning the frequency of the pump-laser modulation around the effective Larmor frequency  expressed by the eq.\eqref{eq:fdiathetaphi}. The resonance line-width is about 25 Hz, however the high S/N ratio let a best fit targeted to a Lorentzian profile determine the peak frequency with a sub-Hz accuracy. 
On the other hand, the ambient field fluctuations (the BB magnetometer is operated in an unshielded environment, with deactivated field-stabilization system \cite{biancalana_prapplml_19}) are at nT level, and this leads to a few Hz uncertainty in the resonant frequency estimation.

The measurements shown here are performed at $B_0=$\SI{1.7}{\micro\tesla} corresponding to an undressed Larmor frequency of \SI{6}{\kilo\hertz}. The dressing field frequency is set at \SI{15}{\kilo\hertz} or \SI{30}{\kilo\hertz}, and its amplitude $B_D$ is set to get a dressing parameter $\xi=1.84$, according to the indication provided by the eq.\eqref{eq:dfdtheta}. In these conditions, the dressed frequency with no tuning field, is $f_0=J_0(1.84) \gamma B_0/2 \pi \approx $ \SI{1900}{\hertz}.

When the tuning field is activated, the measured peak frequency has a remarkable dependence on both the amplitude and the phase of $B_t$ at the sensor location, according to eq.\eqref{eq:fdiathetaphi}. Both $A$ and $\theta$ can be determined by scanning the relative phase of the RF generators in the $[0, 2\pi]$ interval.

The Fig.\ref{fig:sinusoidi} shows clearly that an increase in disk radius causes both a reduction of the amplitude $A$ and a phase shift of the sinusoidal profiles. It is also clear that larger variations of the effective Larmor frequency occur when $\phi=\pm \pi/2$, while operating near the zero-crossings ($\phi=0, \pi$) improves the sensitivity to $\phi$ (i.e to $\theta$) variations, consistently with the eqs.\eqref{eq:dfdA}, \eqref{eq:dfdtheta}.

\begin{figure}
    \centering
    \includegraphics[width=0.9 \columnwidth]{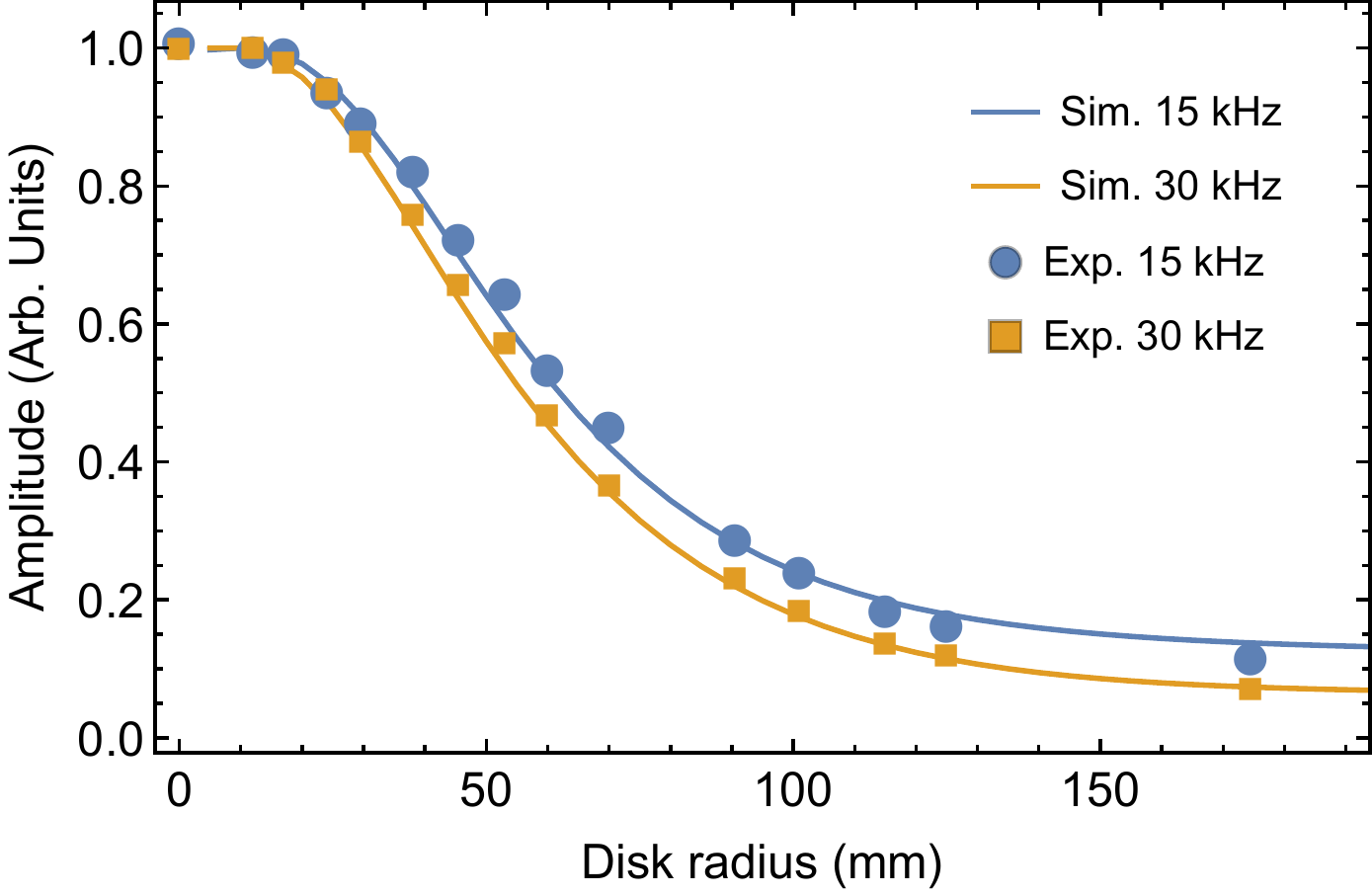}
    \caption{Dependence of the  $B_t$ amplitude on the  Al disk radius
      inferred  from experimental  measurements  (dots) and  evaluated
      from  a  simplified  numerical  simulation (line, see  text  for  more
      details). Both  the cases of $\omega_D= 2\pi\cdot$\SI{15}{kHz} and
      $2\pi \cdot$\SI{30}{kHz} are considered.  The thinner skin depth
      at  \SI{30}{kHz}  produces  a  faster  decay to a lower level, when the disk radius increases,  as expected.  }
    \label{fig:parametriapp2}
\end{figure}

A  clearer  visualization  is  facilitated by  a  best  fit  procedure
targeted to the function  expressed in eq.\ref{eq:fdiathetaphi}. Plots
of the  best-fit parameter $A$ versus  the disk radius are  shown with
dots in  Fig.\ref{fig:parametriapp2}. The  solid lines in  that figure
are results from a finite-elements numerical simulation. The latter is
based on the geometry  of Fig.\ref{fig:interazione}, without the $C_D$
solenoid and considering $B_t$ at the center of the sensor as produced
by a point dipolar source and attenuated by on axis Al disks.

This analysis has been performed with the set of Al disks described in
Sec.\ref{sec:setup} and we report results  obtained at both the tested
tuning-dressing  frequencies.   Selecting   different  values  of  the
oscillating  fields changes  the  skin depth  and  hence modifies  the
shielding  properties of  the Al  samples.  As known,  the skin  depth
depends on  the absolute permeability $\mu$  of the medium and  on its
resistivity $\rho$, according to:
\begin{equation}
\label{eq:pelle}
\delta=\sqrt{\frac{2\rho}{p\omega_D\mu}}
\end{equation}
In the case of Al, this formula gives $\delta=$ \SI{460}{\micro\metre}
and  \SI{650}{\micro\metre},  for   the  two  considered  frequencies,
respectively. Both  values are much  larger than the thickness  of the
foil  from  which  the  Aluminum  disks  (\SI{30}{\micro\metre})  are
cropped.

The   general  trend   of   the  experimental   results  reported   in
Fig.\ref{fig:parametriapp2} is  qualitatively reproduced  by numerical
simulation. In particular both the  experiment and the simulation show
that  the curve  width and  the asymptotic  values decrease  at higher
$\omega_D$ (smaller $\delta$).
However, some  quantitative discrepancies emerge: double  thickness Al
disks are  considered in the  simulation to produce the  well matching
profiles  reported  in  Fig.\ref{fig:parametriapp2}.  Indeed,  several
details are neglected  in the simulation, such as the  presence of the
solenoid $C_D$ and of an electric  heater surrounding the Cs cell, and
the finite  size of the  $B_t$ source and  of the sensor  volume. Very
likely one of more of these factors are  responsible for the mentioned
discrepancies. Similar issues are encountered in the phase estimations
$\theta (r)$.

\section{Discussion}
\label{sec:discussion}

The proposed tuning-dressing technique, extends to all-optical atomic magnetometric sensors  applicability as detectors in  EII setups. Interestingly, the proposed arrangement could be easily implemented in pre-existing  EII apparatuses, and dual modality (RF and tuning-dressing) setups could be built, with incremental complexity of the available instrumentation.

In arrangements where the pump and probe beams are not parallel, a dual-mode operation would make possible to apply the RF field along different orientation,  in such a way to induce variously distributed eddy currents. In fact, while in RF magnetometry the RF field is applied perpendicularly to the pump beam, in the tuning-dressing BB  arrangement $B_t$ is perpendicular to the probe beam.

Compared to conventional RF apparatuses,  the tuning-dressing implementation besides not requiring that the detection field oscillates at a resonant frequency. Moreover, fast selection of different $B_t$ frequencies is feasible, via an appropriate selection of the harmonic parameter $p$. The possibility of selecting arbitrary values for both $\omega$ and $p$ is relevant for the skin depth dependence on $p \omega_D$ (eq. \ref{eq:pelle}), which is of interest to improve the tomographic (three-dimensional) potential of the EII apparatuses.

Concerning the detection efficiency, the proposed methodology takes advantage from the fact that in the tuning-dressed configuration the weak tuning field borrows strength from the primary dressing field. The linear dependence on $B_t$ occurring for odd $p$ (which can be maximized with appropriate choices of $\xi$ and selecting $\phi=\pm \pi/2$) makes possible and efficient the use of approaches as that described in ref.\cite{deans_apl_20}. Specifically it would be possible to apply a vanishing $B_t$ obtained as a superposition of two opposite contributions, in such way that the specimen causes a $B_t$ variation via an unbalance of the two applied terms.

Improved detection techniques could be developed on the basis of the harmonic dependence on $\phi$. As an example, if $B_t$ oscillates at frequency $\omega_t$ slightly different from $p \omega$, the signal will appear as a low-frequency beating term oscillating at $\omega_t-p \omega$: a feature that would enable the application of phase-sensitive detection techniques.

\section{Conclusion}
We have presented a proof of principle experiment proposing a new kind of magnetometric detector of small or weakly conductive specimens. Specifically, we have used a recently studied dressing-field configuration to make an all-optical Bell-and-Bloom magnetometer suited to detect faint oscillating fields as occurring in magnetic induction tomography setups. The experiment demonstrated that the variation of a non-resonantly oscillating probe field can be effectively measured to detect the presence of conductive objects. The described approach can be regarded as non-resonant alternative to the well known methodologies based on RF magnetometers. It constitutes indeed a complementary approach to the problem and dual-mode (RF and tuning-dressing) apparatuses can be envisaged, with interesting novel features and enhanced flexibility.
\label{sec:conclusion}


\bibliographystyle{ieeetr}
\bibliography{bibliograph}

\begin{thebibliography}{10}

\bibitem{griffiths_mst_01}
H.~Griffiths, ``Magnetic induction tomography,'' {\em Measurement Science and
  Technology}, vol.~12, pp.~1126--1131, jul 2001.

\bibitem{deans_apl_20}
C.~Deans, L.~Marmugi, and F.~Renzoni, ``Sub-{Sm}$^{–1}$ electromagnetic
  induction imaging with an unshielded atomic magnetometer,'' {\em Applied
  Physics Letters}, vol.~116, no.~13, p.~133501, 2020.

\bibitem{yu_el_93}
Z.~Z. {Yu}, A.~T. {Peyton}, M.~S. {Beck}, W.~F. {Conway}, and L.~A. {Xu},
  ``Imaging system based on electromagnetic tomography ({EMT}),'' {\em
  Electronics Letters}, vol.~29, no.~7, pp.~625--626, 1993.

\bibitem{Korjenevsky_pm_00}
A.~Korjenevsky, V.~Cherepenin, and S.~Sapetsky, ``Magnetic induction
  tomography: experimental realization,'' {\em Physiological Measurement},
  vol.~21, pp.~89--94, feb 2000.

\bibitem{bevington_as_20}
P.~Bevington, R.~Gartman, and W.~Chalupczak, ``Inductive imaging of the
  concealed defects with radio-frequency atomic magnetometers,'' {\em Applied
  Sciences}, vol.~10, no.~19, 2020.

\bibitem{soleimani_ieee_07}
M.~Soleimani, W.~R.~B. Lionheart, and A.~J. Peyton, ``Image reconstruction for
  high-contrast conductivity imaging in mutual induction tomography for
  industrial applications,'' {\em IEEE Transactions on Instrumentation and
  Measurement}, vol.~56, no.~5, pp.~2024--2032, 2007.

\bibitem{ma_mst_17}
L.~Ma and M.~Soleimani, ``Magnetic induction tomography methods and
  applications: a review,'' {\em Measurement Science and Technology}, vol.~28,
  p.~072001, jun 2017.

\bibitem{deans_apl_16}
C.~Deans, L.~Marmugi, S.~Hussain, and F.~Renzoni, ``Electromagnetic induction
  imaging with a radio-frequency atomic magnetometer,'' {\em Applied Physics
  Letters}, vol.~108, no.~10, p.~103503, 2016.

\bibitem{wickenbrock_apl_16}
A.~Wickenbrock, N.~Leefer, J.~W. Blanchard, and D.~Budker, ``Eddy current
  imaging with an atomic radio-frequency magnetometer,'' {\em Applied Physics
  Letters}, vol.~108, no.~18, p.~183507, 2016.

\bibitem{jensen_prr_19}
K.~Jensen, M.~Zugenmaier, J.~Arnbak, H.~St\ae{}rkind, M.~V. Balabas, and E.~S.
  Polzik, ``Detection of low-conductivity objects using eddy current
  measurements with an optical magnetometer,'' {\em Phys. Rev. Research},
  vol.~1, p.~033087, Nov 2019.

\bibitem{marmugi_apl_19}
L.~Marmugi, C.~Deans, and F.~Renzoni, ``Electromagnetic induction imaging with
  atomic magnetometers: Unlocking the low-conductivity regime,'' {\em Applied
  Physics Letters}, vol.~115, no.~8, p.~083503, 2019.

\bibitem{scharfetter_ieee_03}
H.~{Scharfetter}, R.~{Casanas}, and J.~{Rosell}, ``Biological tissue
  characterization by magnetic induction spectroscopy (mis): requirements and
  limitations,'' {\em IEEE Transactions on Biomedical Engineering}, vol.~50,
  no.~7, pp.~870--880, 2003.

\bibitem{biancalana_arx_21}
G.~Bevilacqua, V.~Biancalana, Y.~Dancheva, A.~Fregosi, and A.~Vigilante, ``Spin
  dynamic response to a time dependent field,'' 2021.

\bibitem{haroche_prl_1970}
S.~Haroche, C.~Cohen-Tannoudji, C.~Audoin, and J.~P. Schermann, ``Modified
  {Zeeman} hyperfine spectra observed in $^1${H} and $^{87}${Rb} ground states
  interacting with a nonresonant {RF} field,'' {\em Phys. Rev. Lett.}, vol.~24,
  pp.~861--864, 1970.

\bibitem{bevilacqua_prl_20}
G.~Bevilacqua, V.~Biancalana, A.~Vigilante, T.~Zanon-Willette, and E.~Arimondo,
  ``Harmonic fine tuning and triaxial spatial anisotropy of dressed atomic
  spins,'' {\em Phys. Rev. Lett.}, vol.~125, p.~093203, Aug 2020.

\bibitem{biancalana_apb_16}
G.~Bevilacqua, V.~Biancalana, P.~Chessa, and Y.~Dancheva, ``Multichannel
  optical atomic magnetometer operating in unshielded environment,'' {\em
  Applied Physics B}, vol.~122, no.~4, p.~103, 2016.

\bibitem{dodd_jap_68}
C.~V. Dodd and W.~E. Deeds, ``Analytical solutions to eddy‐current
  probe‐coil problems,'' {\em Journal of Applied Physics}, vol.~39, no.~6,
  pp.~2829--2838, 1968.

\bibitem{biancalana_prapplml_19}
G.~Bevilacqua, V.~Biancalana, Y.~Dancheva, and A.~Vigilante, ``Self-adaptive
  loop for external-disturbance reduction in a differential measurement
  setup,'' {\em Phys. Rev. Applied}, vol.~11, p.~014029, Jan 2019.

\end{thebibliography}

\end{document}